\documentclass[a4paper,11pt]{article}
\usepackage{pos}
\usepackage[absolute]{textpos}

\usepackage{siunitx}
\renewcommand{\Re}{\text{Re}\,}
\renewcommand{\Im}{\text{Im}\,}
\newcommand{\hc}{\,\text{h.c.}}
\newcommand{\s}{\par}
\renewcommand{\logo}{\relax} %remove PoS logo for arXiv submission

\AtBeginDocument{\begin{textblock*}{3cm}[1,0](19.3cm,5cm)\raggedleft\texttt{\small KA-TP-01-2023}\end{textblock*}}

\title{`CP in the Dark' and a Strong First-Order Electroweak Phase Transition}
\ShortTitle{`CP in the Dark' and an SFOEWPT}

\author*[a]{Lisa Biermann}
\author[a]{Margarete M\"uhlleitner}
\author[a]{Jonas M\"uller}

\affiliation[a]{Institute for Theoretical Physics, Karlsruhe Institute of Technology,\\
76128 Karlsruhe, Germany}

\emailAdd{lisa.biermann@kit.edu}
\emailAdd{margarete.muehlleitner@kit.edu}

\abstract{We investigate the potential of the model `CP in the Dark' for providing a strong first-order electroweak phase transition (SFOEWPT) by taking into account all relevant theoretical and experimental constraints. 
For the derivation of the strength of the phase transition we use the one-loop corrected, daisy-resummed effective potential at finite temperature, implemented in the C++ code {\tt BSMPT}, to determine the global minimum at the critical temperature. 
The model `CP in the Dark' provides a dark matter (DM) candidate as well as explicit CP violation in the dark sector. We find a broad range of viable parameter points providing an SFOEWPT. 
They are within the reach of XENON1T and future invisible Higgs decay searches for DM. 
`CP in the Dark' also offers SFOEWPT points that feature spontaneous CP violation at finite temperature. 
Having not only an SFOEWPT that provides the necessary departure from thermal equilibrium, but also a source of additional non-standard CP violation, opens a promising gate towards enabling the generation of the baryon asymmetry of the universe (BAU) through electroweak baryogenesis.}

\FullConference{%
  8th Symposium on Prospects in the Physics of Discrete Symmetries (DISCRETE 2022)\\
  7-11 November, 2022\\
  Baden-Baden, Germany
}

\begin{document}
\maketitle

\section{Introduction}
Electroweak baryogenesis (EWBG)~\cite{Kuzmin:1985mm,Cohen:1990it} can generate the observed baryon asymmetry of the universe (BAU) $\eta\simeq\num{6.1e-10}$~\cite{Planck:2018vyg} through bubble formation in the early universe if the Sakharov conditions~\cite{Sakharov:1967dj} are fulfilled.
While baryon-number violating processes can be mediated through electroweak sphalerons at finite temperature in the early universe~\cite{Manton:1983nd,Klinkhamer:1984di} already in the Standard Model (SM) of particle physics, the amount of CP violation and the strength of the departure from thermal equilibrium are insufficient~\cite{Gavela:1993ts,Kajantie:1996mn,Csikor:1998eu}.
Therefore, we need physics beyond the SM (BSM) to first generate enough left-handed fermion access that can then create the BAU in front of the bubble wall through electroweak sphaleron transitions and second, conserve the BAU through sufficiently suppressing the baryon number violating sphaleron transitions in the broken vacuum inside the bubbles.
The latter can be realized in an EWBG scenario through a strong first-order electroweak phase transition (SFOEWPT), when $\xi_c\equiv\frac{v_c}{T_c}\gtrsim 1$, so if the electroweak vacuum expectation value (VEV) $v_c$ at the critical temperature $T_c$ is greater than the critical temperature itself.
The critical temperature $T_c$ marks the temperature where the symmetric and the broken minimum of the effective potential are degenerate.
Here, we discuss an extended scalar sector model candidate called `CP in the Dark' which allows for explicit CP violation in its dark sector.
The model is introduced in Sec.~\ref{sec:model} and its potential to allow for an SFOEWPT as well as phenomenological consequences are discussed in Sec.~\ref{sec:results}, for more details we refer to~\cite{Biermann:2022meg}.

\section{`CP in the Dark'}\label{sec:model}
The model `CP in the Dark' has a Next-to-Minimal 2-Higgs-Doublet Model (N2HDM)-like extended scalar sector where the additional complex $SU(2)_L$-doublet $\Phi_2$ and the additional real $SU(2)_L$-singlet $\Phi_S$ both transform odd under \textit{one} additionally implied $\mathbb{Z}_2$-symmetry \cite{Azevedo:2018fmj}
\begin{align}
  \Phi_1\rightarrow\Phi_1\,,\quad\Phi_2\rightarrow-\Phi_2\,,\quad\Phi_S\rightarrow-\Phi_S\,.\label{eq:Z2Symmetry}
\end{align}
As the Yukawa sector is required to be neutral under this discrete symmetry, only $\Phi_1$ couples to fermions.
Therefore, the absence of scalar-mediated tree-level flavour-changing neutral currents (FCNCs) is ensured and the Yukawa sector is identical to the SM one.
The most general invariant scalar tree-level potential reads
\begin{equation}
  \begin{split}
    V^{(0)}\,=\,\,\,&m_{11}^2|\Phi_1|^2+m_{22}^2|\Phi_2|^2+\frac{m_S^2}{2}\Phi_S^2+\left( A \Phi_1^\dagger \Phi_2 \Phi_S + \hc \right)\\
    &+\frac{\lambda_1}{2}|\Phi_1|^4+\frac{\lambda_2}{2}|\Phi_2|^4+\lambda_3 |\Phi_1|^2|\Phi_2|^2+\lambda_4 |\Phi_1^\dagger \Phi_2|^2+\frac{\lambda_5}{2}[(\Phi_1^\dagger \Phi_2)^2+(\Phi_2^\dagger \Phi_1)^2]\\
    &+\frac{\lambda_6}{4}\Phi_S^4+\frac{\lambda_7}{2}|\Phi_1|^2 \Phi_S^2+\frac{\lambda_8}{2}|\Phi_2|^2\Phi_S^2\,.\label{eq:TreeLevelPotential}
  \end{split}
\end{equation}
The parameters of the potential are real, except for the trilinear coupling $A$. A non-zero imaginary phase of $A$ directly corresponds to explicit CP violation, see below.\s
After electroweak symmetry breaking (EWSB), we expand around the vacuum expectation values (VEVs) and allow for the most general vacuum structure at finite temperature with five VEVs $\omega_i$ ($i\in\{\text{CB},1,2,\text{CP},S\}$)
\begin{align}
  \Phi_1=\frac{1}{\sqrt{2}}\begin{pmatrix}\rho_1+i\eta_1\\\zeta_1+\omega_1+i\Psi_1\end{pmatrix}\,,\quad
  \Phi_2=\frac{1}{\sqrt{2}}\begin{pmatrix}\rho_2+\omega_\text{CB}+i\eta_2\\\zeta_2+ \omega_2+i(\Psi_2+\omega_\text{CP})\end{pmatrix}\,,\quad
  \Phi_S=\zeta_S+\omega_S\,.\label{eq:VacuumTFinite}
\end{align}
The zero-temperature vacuum is chosen to conserve the $\mathbb{Z}_2$-symmetry of Eq.~\ref{eq:Z2Symmetry},
\begin{align}
  \langle\Phi_1\rangle|_{T=\,\SI{0}{GeV}}=\frac{1}{\sqrt{2}}\begin{pmatrix}0\\v_1\end{pmatrix}\,,\quad
  \langle\Phi_2\rangle|_{T=\,\SI{0}{GeV}}=\begin{pmatrix}0\\0\end{pmatrix}\,,\quad
  \langle\Phi_S\rangle|_{T=\,\SI{0}{GeV}}=0\,,\label{eq:VacuumTZero}
\end{align}
with $\overline{\omega}_1 |_{T=\,\SI{0}{GeV}}\equiv v_1 \equiv v\approx\SI{246.22}{GeV}$ fixing $\Phi_1$ to be the SM-like doublet providing the SM-like Higgs boson $h$ as its neutral CP-even mass eigenstate.
In the following, we choose the notation $\overline{\omega}_i$ for the coordinate of $\omega_i$ at the global minimum of the effective potential.
The conservation of the $\mathbb{Z}_2$-symmetry at $T=\SI{0}{GeV}$ results in a new conserved quantum number, called \textit{dark charge}.
SM-like particles originating from the first doublet have dark charge $+1$.
Physical particles originating from the second doublet and the real singlet have dark charge $-1$ and are called \textit{dark particles}.
The dark sector of the model consists of a pair of charged dark scalars $H^\pm$ and three neutral dark scalars $h_1$, $h_2$ and $h_3$ stemming from the mixing of the neutral fields of $\Phi_2$ and $\Phi_S$, with $h_1$ being the lightest and therefore stable particle dark matter (DM) candidate.
The model also features explicit CP violation, introduced through $\Im(A)\neq 0$, which exists solely in the dark sector, being therefore not constrained by electric dipole moment (EDM) constraints \cite{Azevedo:2018fmj, Biermann:2022meg}.
In the following, we will study the potential of this model to provide viable parameter sets that feature an SFOEWPT.

\section{Numerical Results}\label{sec:results}
\begin{figure}[t]
  \includegraphics[width=.5\textwidth]{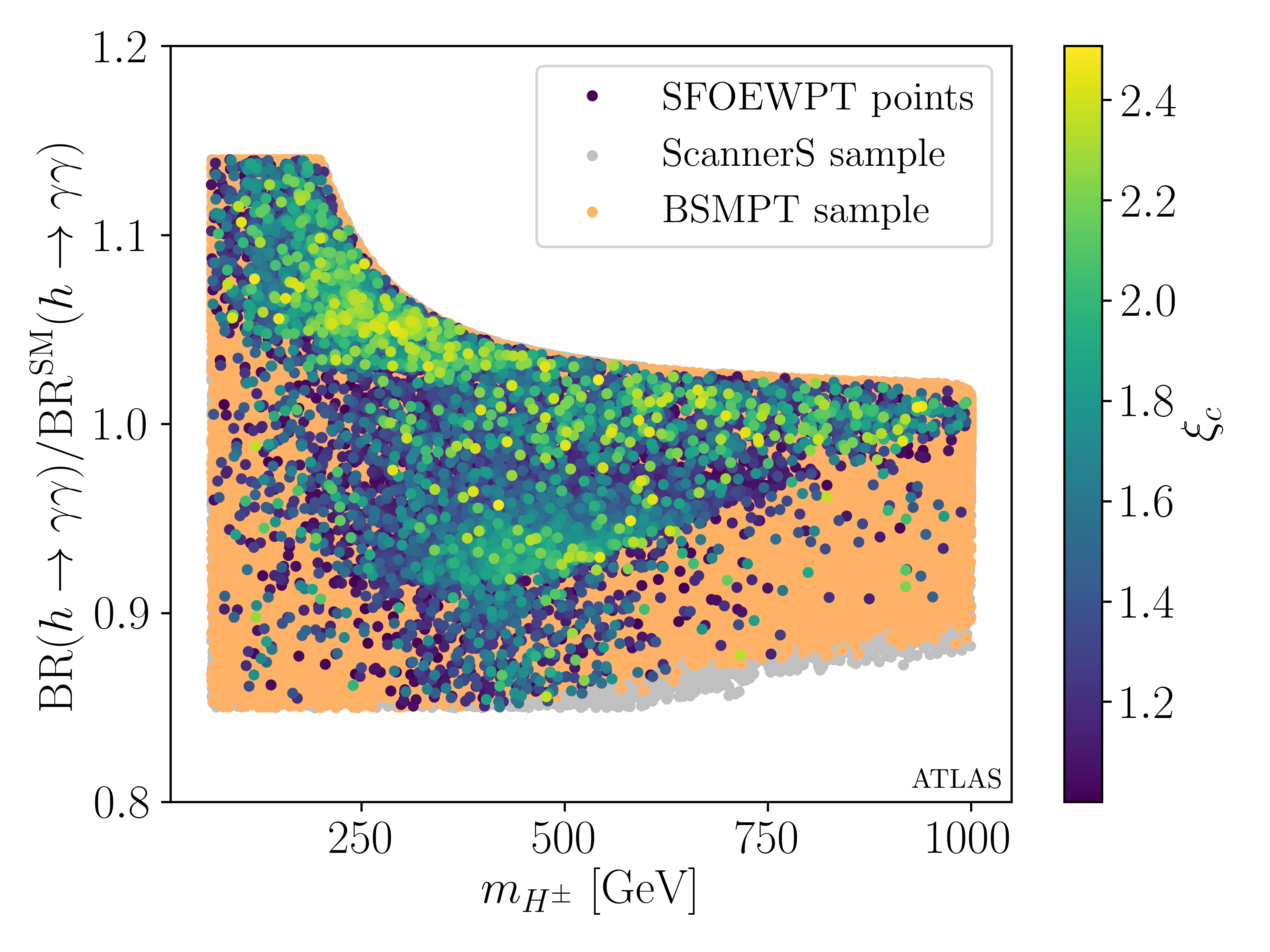}
  \includegraphics[width=.5\textwidth]{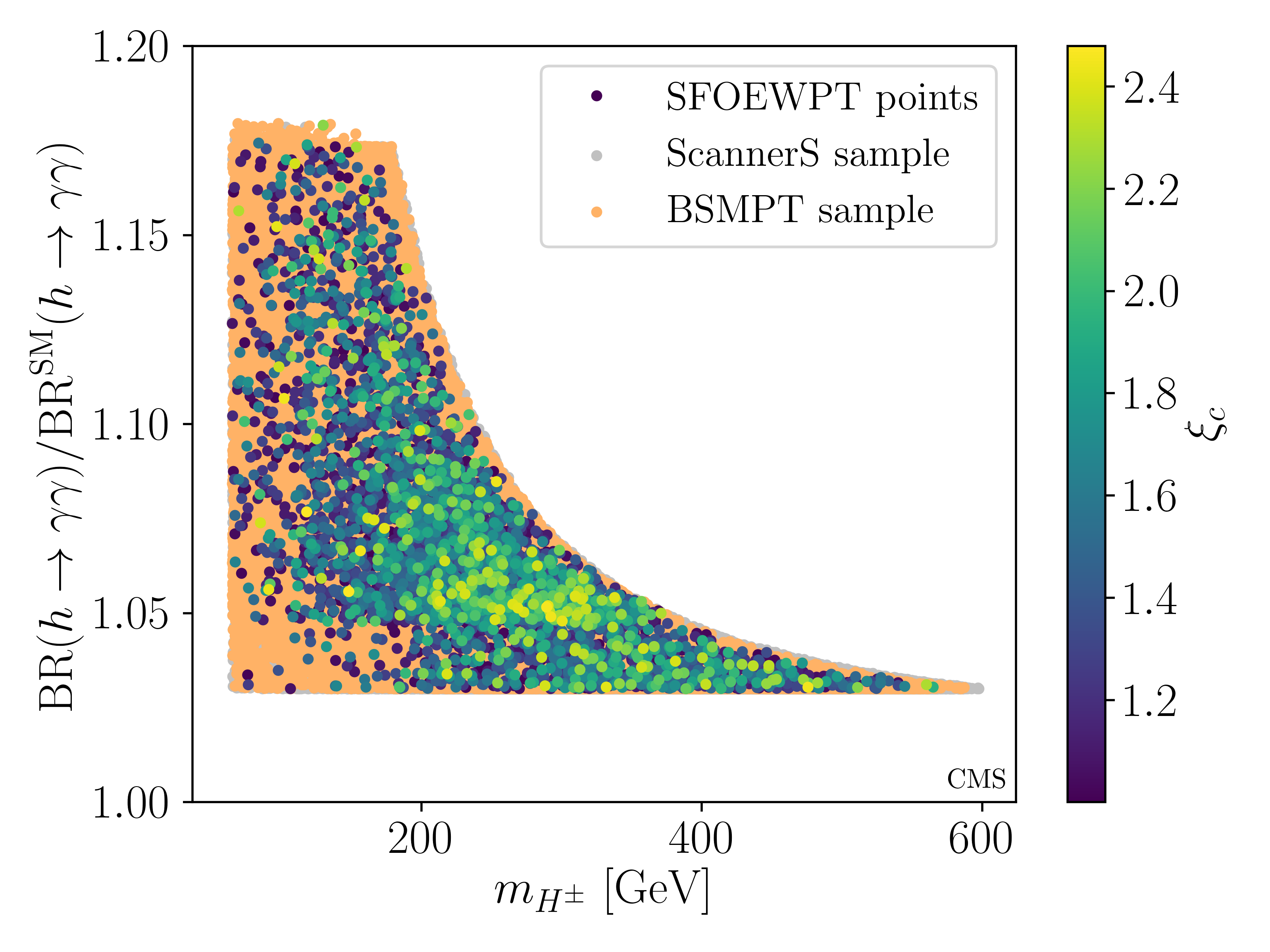}
  \caption{
    Branching ratio of the SM-like Higgs $h$ into a photon pair normalized to its SM value versus the charged dark scalar mass $m_{H^\pm}$ applying the latest limit on $\mu_{\gamma\gamma}$ by {\tt ATLAS}~\cite{ATLAS:2018hxb} (left) $\mu^\text{ATLAS}_{\gamma\gamma} = 0.99^{+0.15}_{-0.14}$ as well as by {\tt CMS}~\cite{CMS:2021kom} (right) $\mu^\text{CMS}_{\gamma\gamma} = 1.12^{\pm 0.09}$.
    The {\tt ScannerS} sample is shown in grey. In orange, we illustrate the {\tt BSMPT} sample. The coloured points are SFOEWPT points.
  }
  \label{fig:br_hsm_gamgam}
\end{figure}

\begin{figure}[t]
  \includegraphics[width=.5\textwidth]{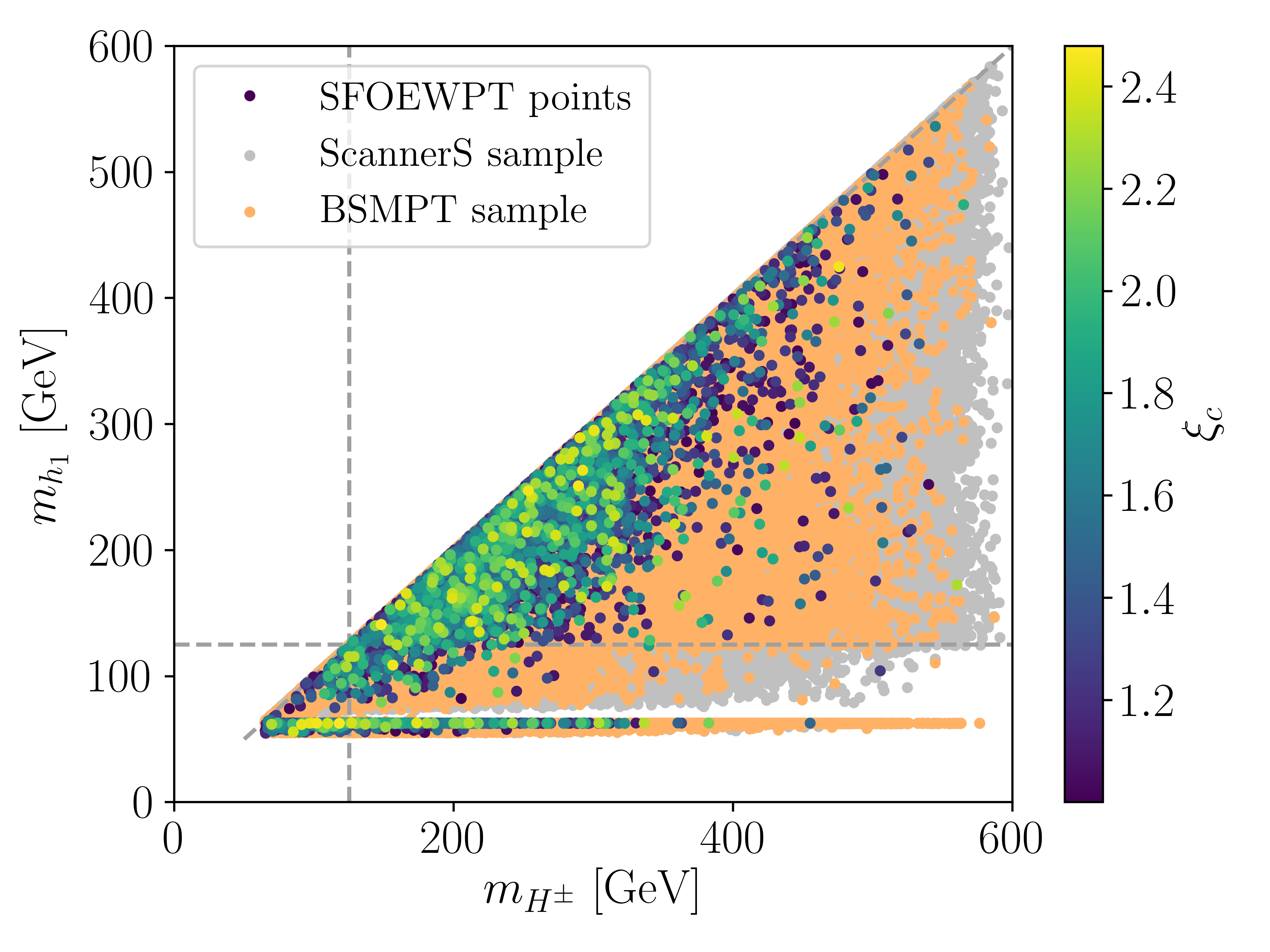}
  \includegraphics[width=.5\textwidth]{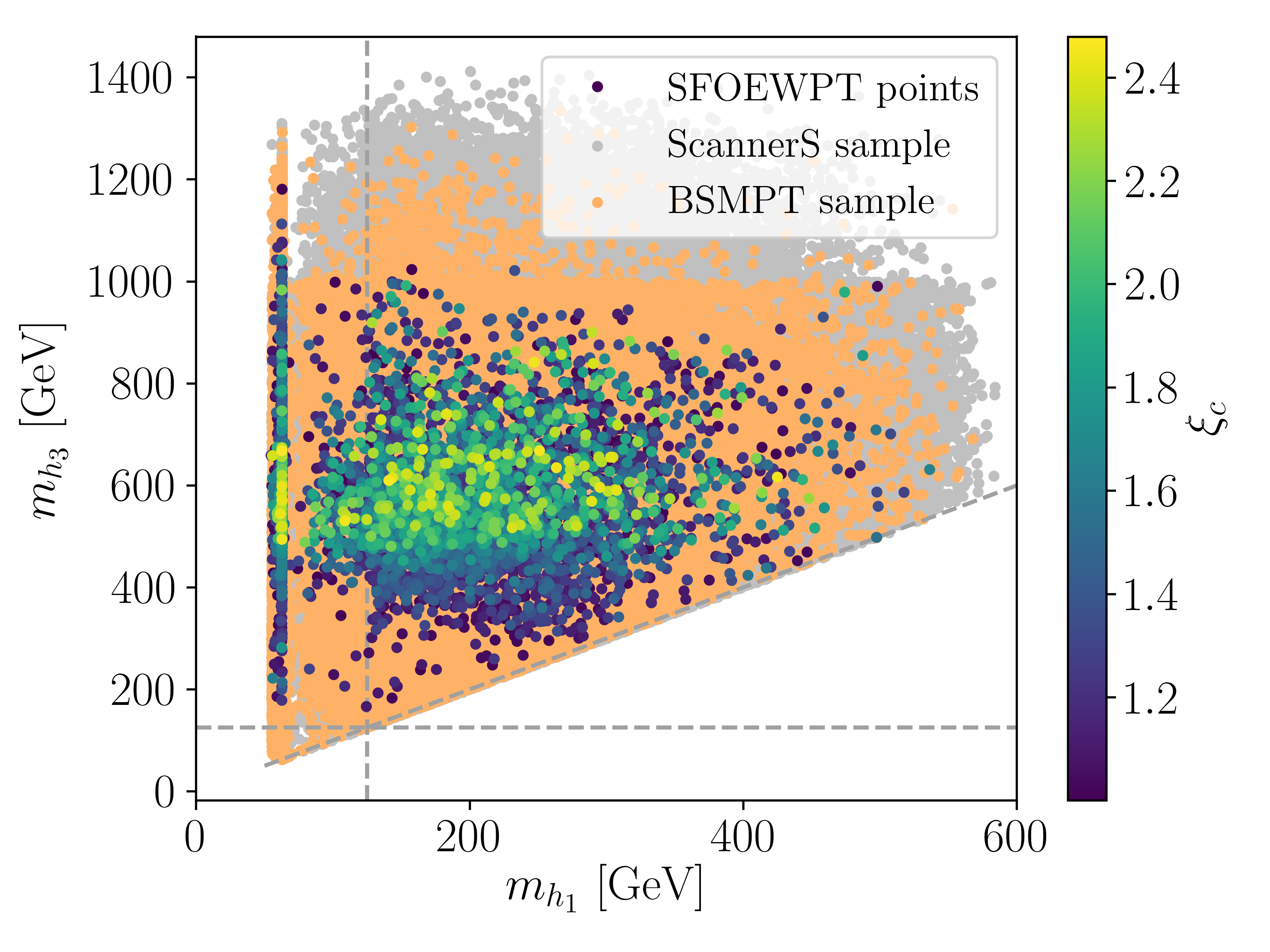}
  \caption{
    Sample point distribution in the mass parameter space spanned by the charged dark scalar mass $m_{H^\pm}$ and the DM mass $m_{h_1}$ (left) as well as the DM mass $m_{h_1}$ and the heaviest neutral dark scalar mass $m_{h_3}$ (right).
    Colour code is the same as in Fig.~\ref{fig:br_hsm_gamgam}.
  }
  \label{fig:mass_distributions_sfoewpt}
\end{figure}

\begin{figure}[h]
  \includegraphics[width=.5\textwidth]{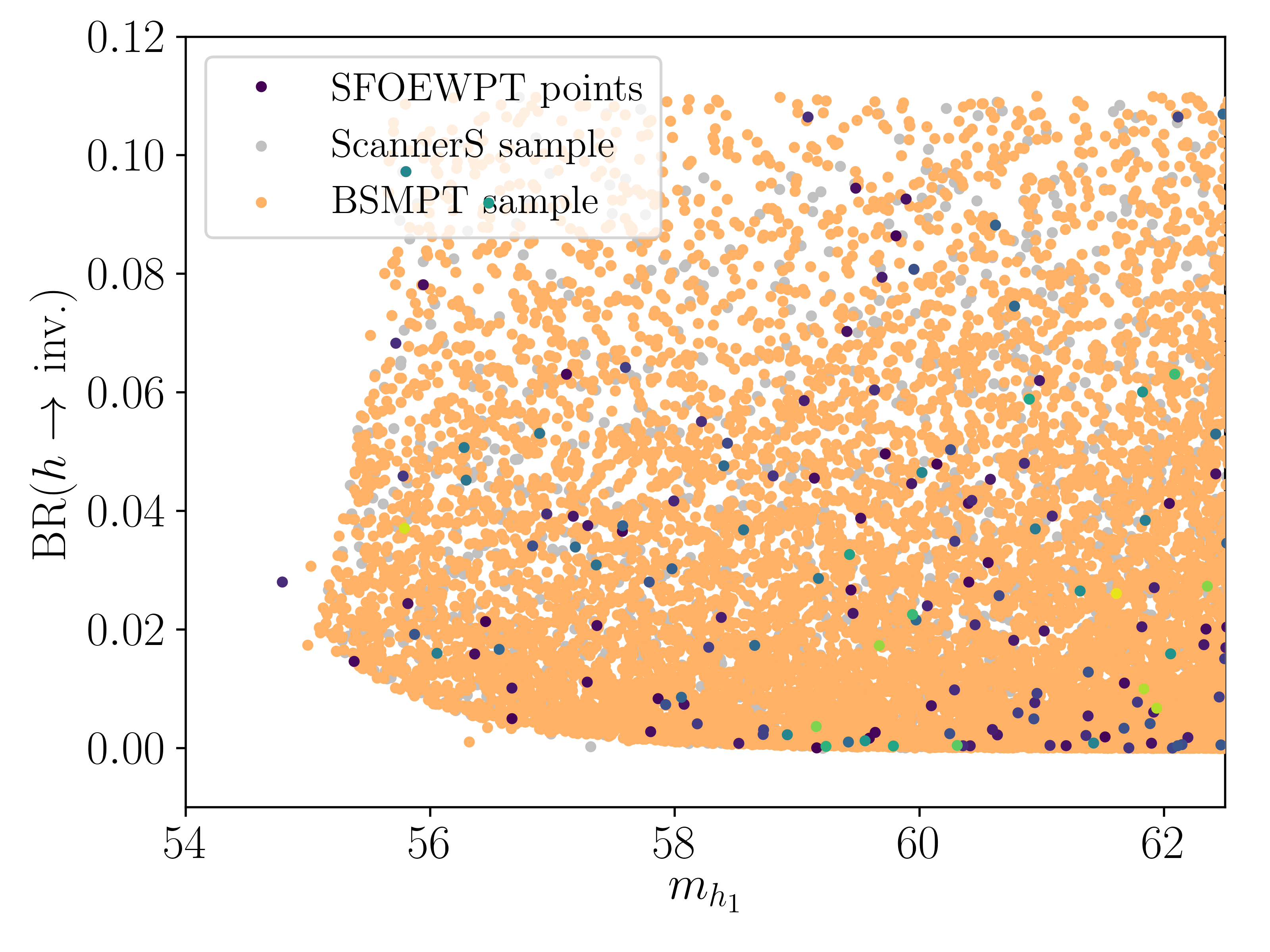}
  \includegraphics[width=.5\textwidth]{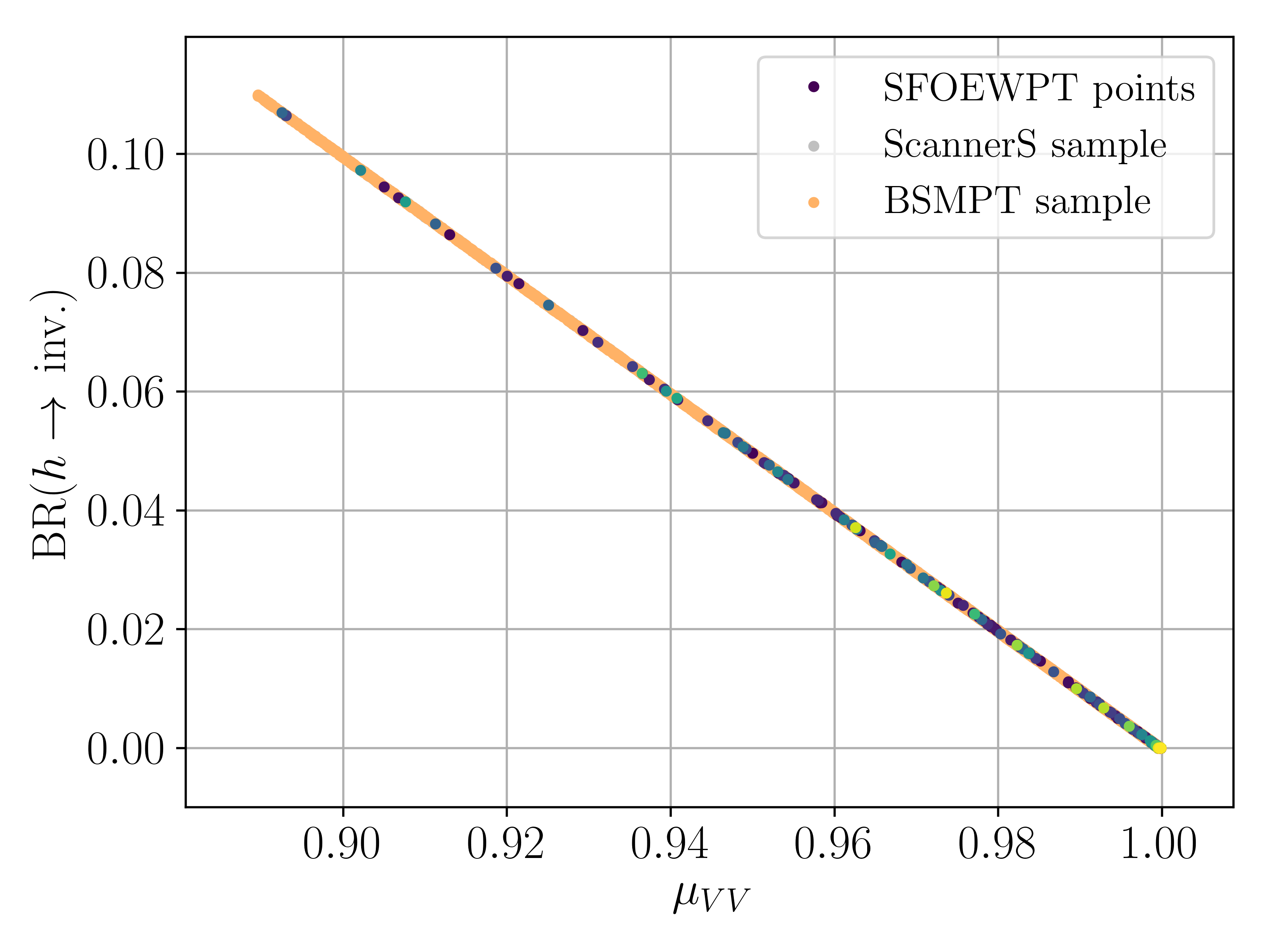}
  \caption{
    Branching ratio of the SM-like Higgs $h$ into invisible particles versus the DM mass $m_{h_1}$ (left) and the gauge boson signal strength $\mu_{VV}$ (right).
    Colour code is the same as in Fig.~\ref{fig:br_hsm_gamgam}.
  }
  \label{fig:br_hsm_inv}
\end{figure}

Our parameter sample is obtained using the code {\tt ScannerS}~\cite{Coimbra:2013qq,Muhlleitner:2020wwk} which tests for the relevant theoretical and experimental constraints, all details can be found in~\cite{Biermann:2022meg}.
On top of these constraints we also require compatibility with the recent ATLAS result on the Higgs decay into invisible particles, $\text{BR}(h\rightarrow\,\text{inv.})<\num{0.11}$~\cite{ATLAS:2020kdi}.
Performing a global minimization of the one-loop corrected, daisy-resummed effective potential at finite temperature in a temperature range of $T\in\{0,300\}\,\si{GeV}$, we determine parameter points of `CP in the Dark' that allow for an SFOEWPT through the usage of a bisection method, as implemented in the code {\tt BSMPT}~\cite{Basler:2018cwe,Basler:2020nrq}.
{\tt BSMPT}, among other things, ensures that the NLO electroweak vacuum is the global minimum of the effective potential at $T=\SI{0}{GeV}$.\s
Figures~\ref{fig:br_hsm_gamgam}-\ref{fig:br_hsm_inv} show our parameter sample obtained by {\tt ScannerS} (in grey), overlaid by the points that in addition have a stable NLO EW vacuum as tested by {\tt BSMPT} (in orange) and overlaid by the found SFOEWPT points with the value of $\xi_c$ indicated by the colour.\s
In Fig.~\ref{fig:br_hsm_gamgam} we show our sample in the plane spanned by the charged dark scalar mass $m_{H^\pm}$ and the branching ratio of the SM-like Higgs $h$ into a photon pair normalized to its SM value after applying the latest limit on $\mu_{\gamma\gamma}$ by {\tt ATLAS}~\cite{ATLAS:2018hxb} (left) $\mu^\text{ATLAS}_{\gamma\gamma} = 0.99^{+0.15}_{-0.14}$ as well as by {\tt CMS}~\cite{CMS:2021kom} (right) $\mu^\text{CMS}_{\gamma\gamma} = 1.12^{\pm 0.09}$. For all following figures, we apply the {\tt CMS} cut noting that a future increased precision on $\mu_{\gamma\gamma}$ has the potential to substantially further cut the parameter space on $m_{H^\pm}$ due to its characteristic shape with an \textit{only} dark sector-induced increase in $\text{BR}(h\to \gamma\gamma)/\text{BR}^\text{SM}(h\to \gamma\gamma)$ towards smaller $m_{H^\pm}$ governed by the quartic coupling $\lambda_3$.\s
In Fig.~\ref{fig:mass_distributions_sfoewpt} we illustrate our parameter sample in the parameter space spanned by the DM mass $m_{h_1}$, as well as charged dark scalar mass $m_{H^\pm}$ and the heaviest neutral scalar mass $m_{h_3}$, respectively. Our parameter space is neither constrained by the requirement of NLO-VEV stability, nor by ensuring an SFOEWPT, as our found viable SFOEWPT points distribute across all the allowed parameter space. The cut on a maximally allowed upper $m^\text{max}_{H^\pm}\approx\SI{597}{GeV}$ stems from the cut on $\mu_{\gamma\gamma}$, as elaborated above and illustrated in Fig.~\ref{fig:br_hsm_gamgam}.

We also find viable SFOEWPT points scattered across all {\tt ScannerS}-allowed parameter space in the low DM-mass range $m_{h_1}<\SI{62.5}{GeV}$, as illustrated in Fig.~\ref{fig:br_hsm_inv} (left), where we show the branching ratio of the SM-like Higgs $h$ into invisible particles versus $m_{h_1}$.
The investigation of the correlation between the branching ratio of the SM-like Higgs into invisible particles and the gauge boson signal strength $\mu_{VV}$ ($V=Z,W$) shows that our results are in agreement with~\cite{Engeln:2020fld}.\s

\begin{figure}[h]
  \includegraphics[width=.5\textwidth]{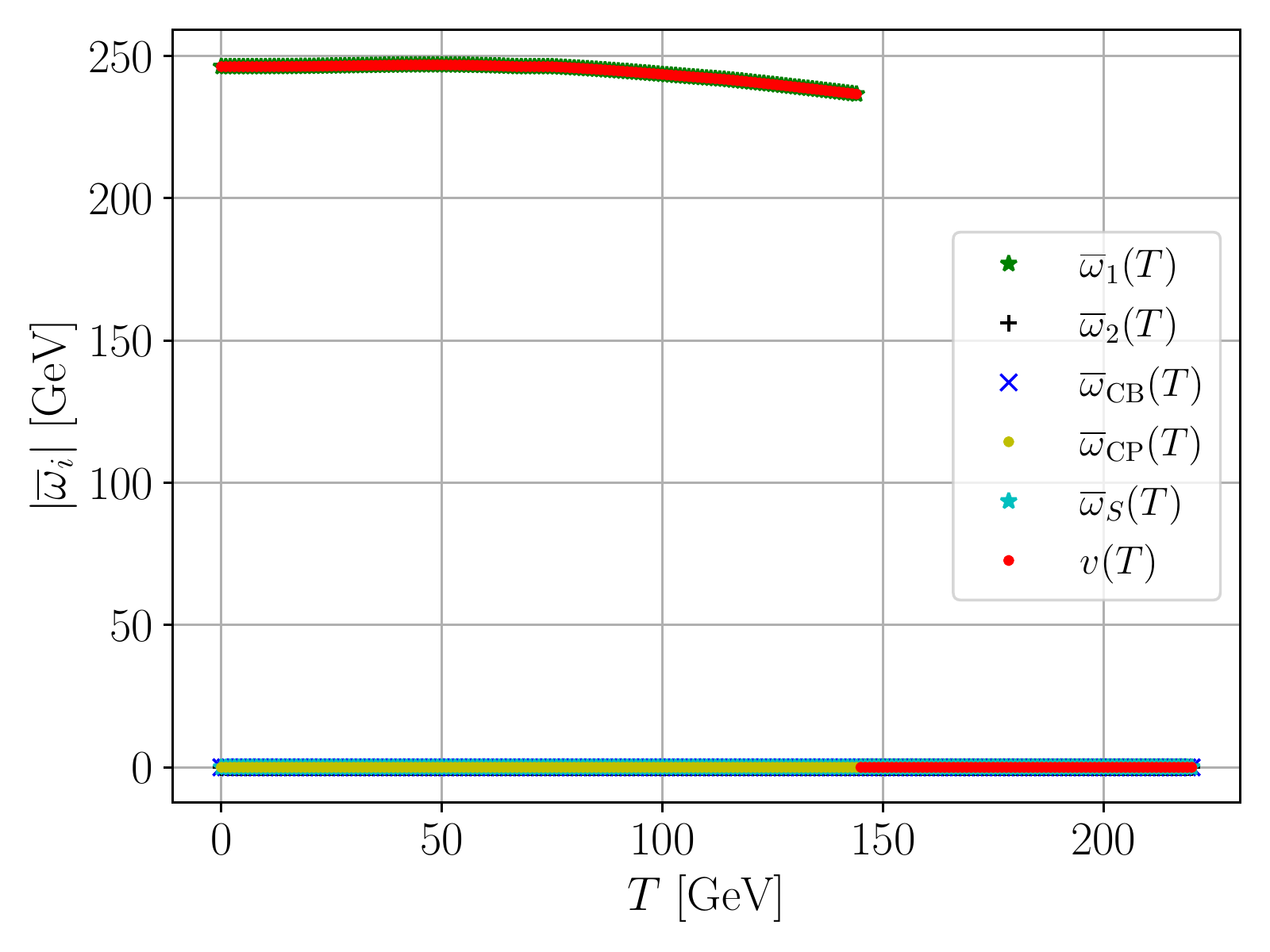}
  \includegraphics[width=.5\textwidth]{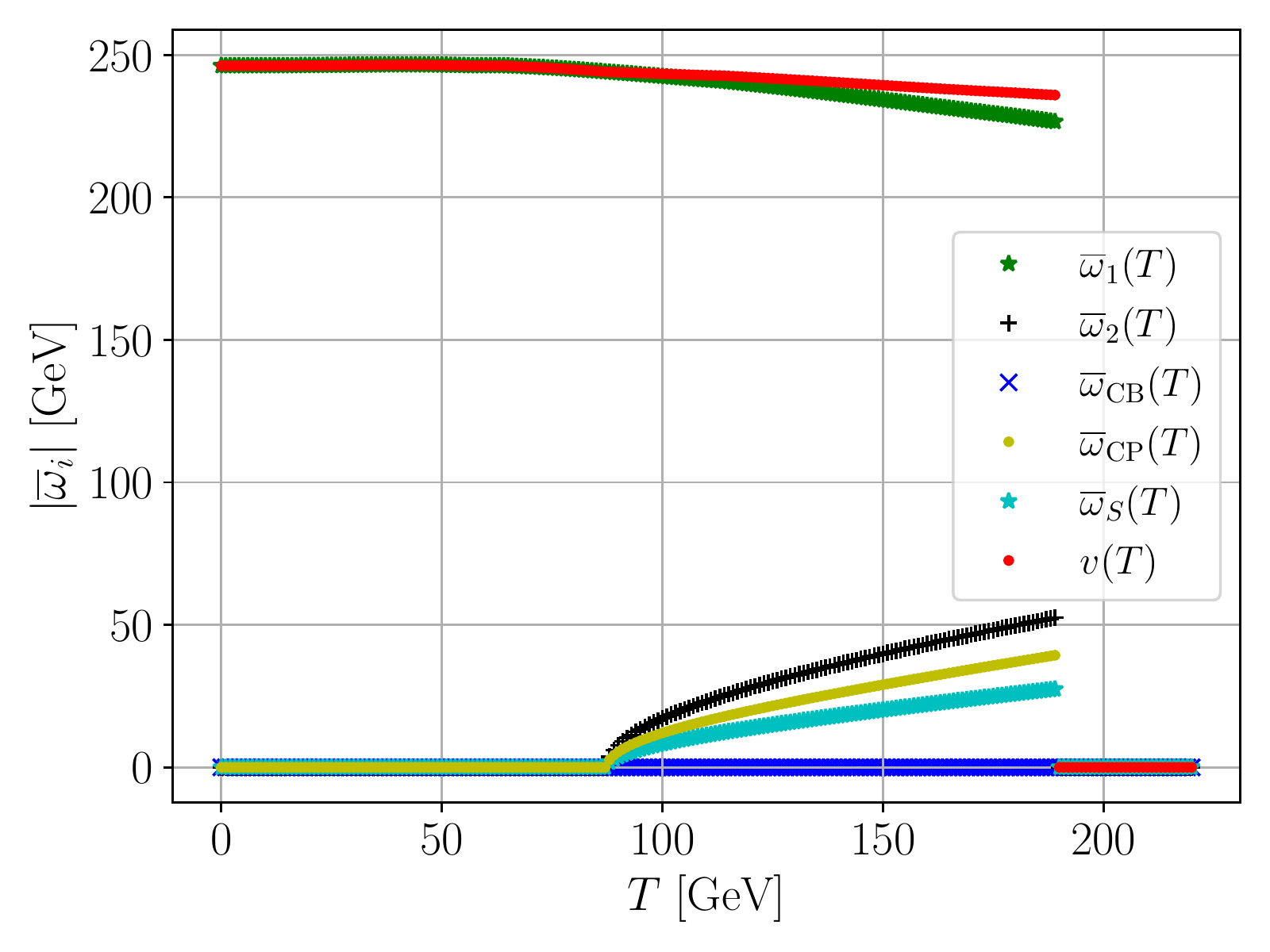}
  \caption{
      Evolution of all VEV direction that minimize the one-loop, daisy-corrected, effective potential at finite temperature for two points with an SFOEWPT. The electroweak VEV is shown in red, on the left for a point where only the SM-like VEV $\overline{\omega}_1$ participates in the phase transition, on the right for a point where the dark VEVs except $\overline{\omega}_\text{CB}$ participate in the phase transition. The benchmark points are explicitly given in~\cite{Biermann:2022meg}.
  }
  \label{fig:vevevo_cpv_ft}
\end{figure}
\begin{figure}[h]
  \centering
  \includegraphics[width=.5\textwidth]{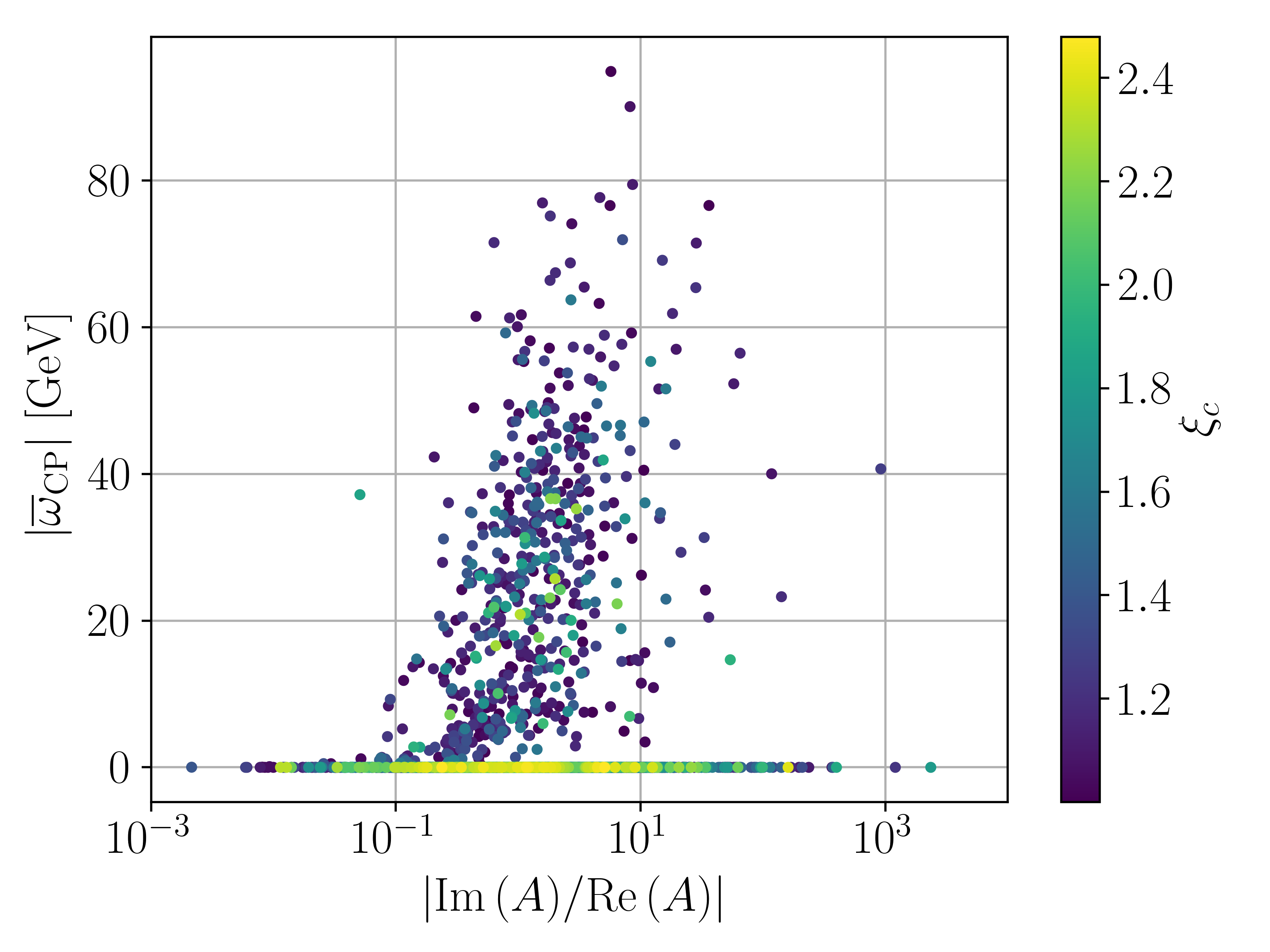}
  \caption{
    Value of the CP-violating VEV $\overline{\omega}_\text{CP}$ versus $|\Im (A)/\Re (A)|$ for all SFOEWPT points.
    The colour indicates the strength $\xi_c$ of the SFOEWPT.
  }
  \label{fig:spon_cpv_ft}
\end{figure}

\begin{figure}[h]
  \includegraphics[width=.5\textwidth]{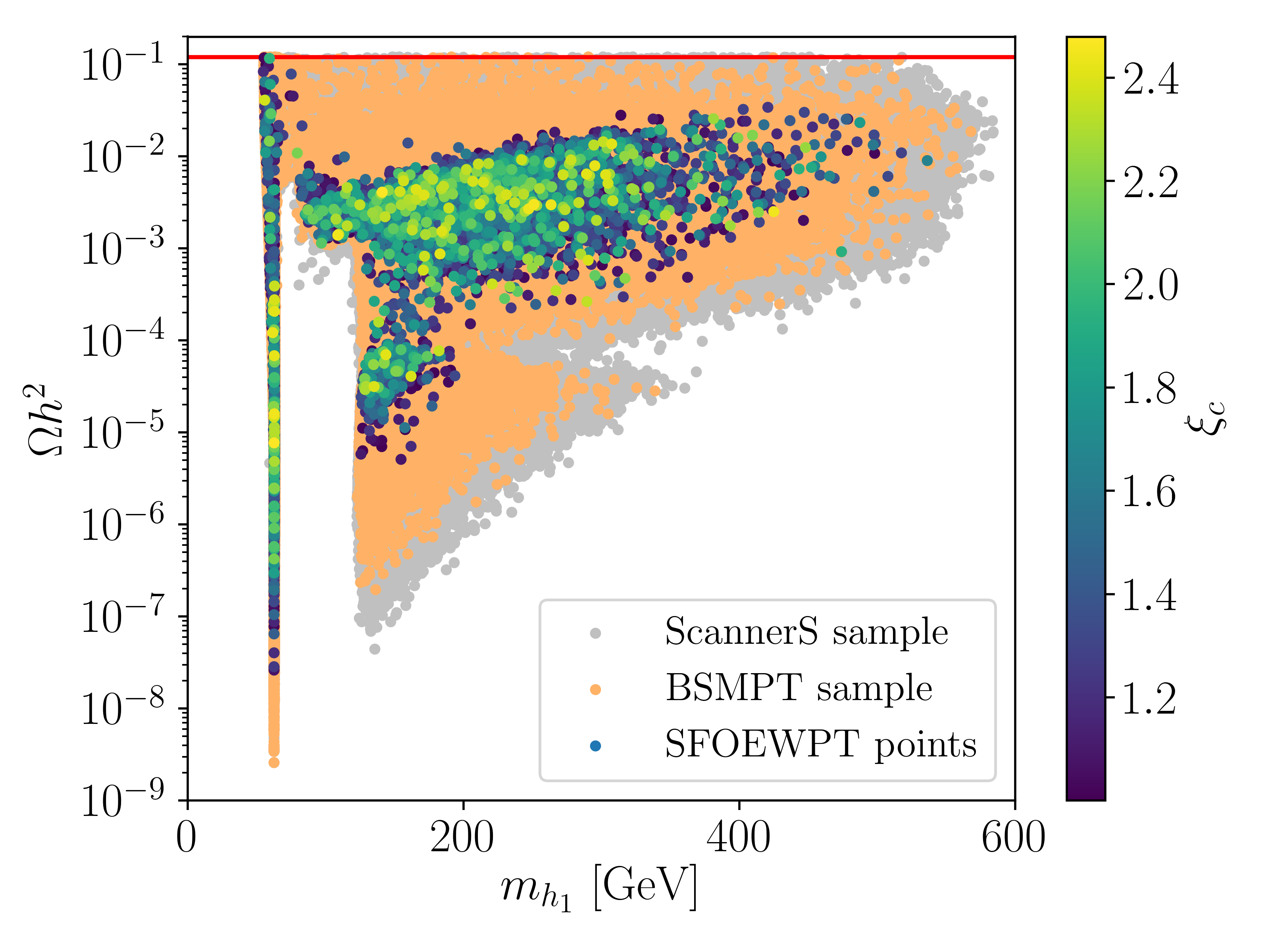}
  \includegraphics[width=.5\textwidth]{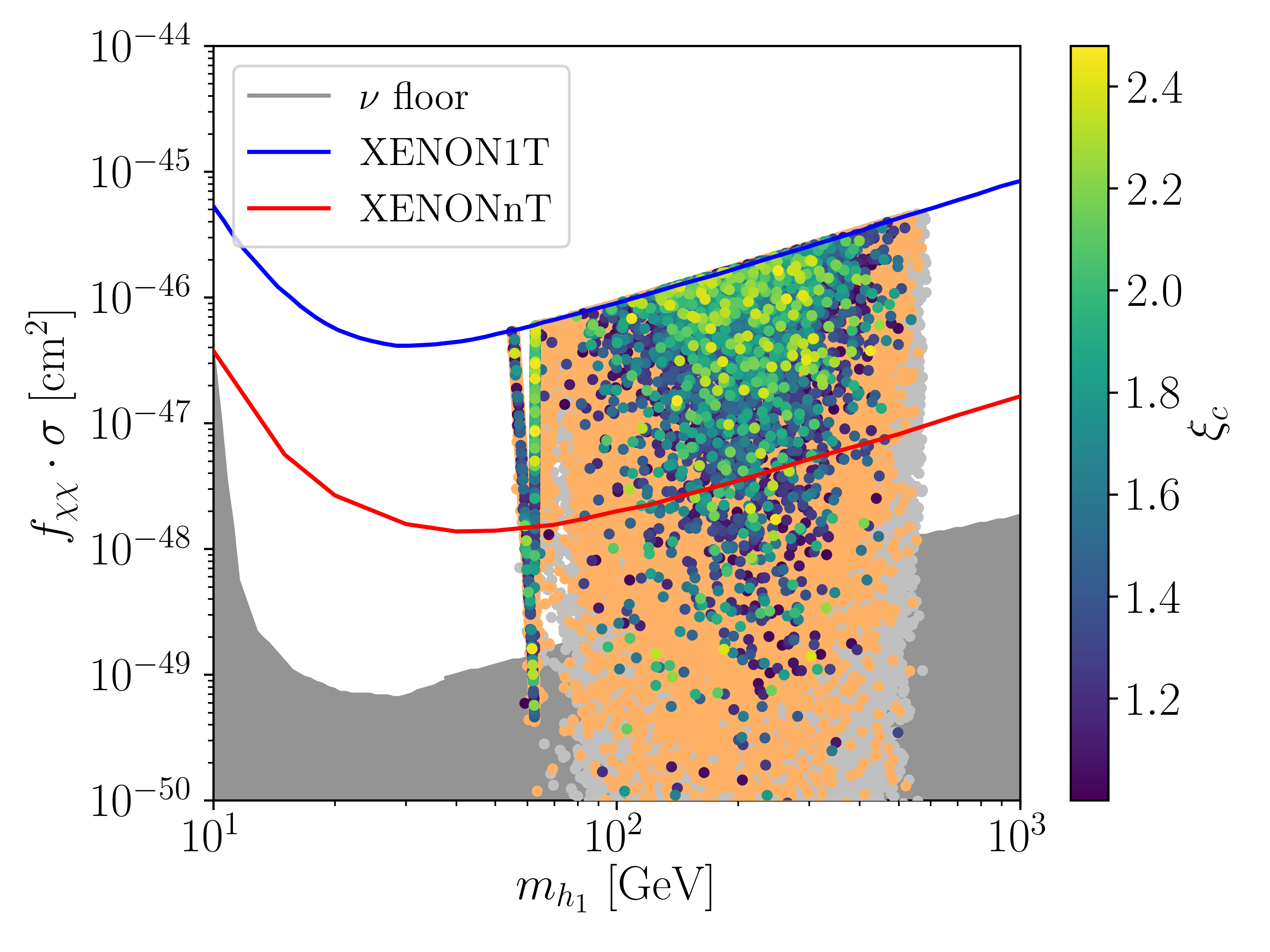}
  \caption{
    Relic density (left) and effective SI nucleon DM cross section (right) versus DM mass $m_{h_1}$.
    The colour code is the same as in Fig.~\ref{fig:br_hsm_gamgam}.
    On the left, we show the experimentally measured relic density $\Omega_\text{obs} h^2 = \num{0.1200\pm0.0012}$~\cite{Planck:2018vyg} in red.
    On the right, we additionally display the {\tt XENON1T} exclusion limit in blue~\cite{XENON:2018voc} and the projected sensitivity of the {\tt XENONnT} experiment in red~\cite{XENON:2020kmp}.
    The experimental limit for the neutrino background (in grey) has been taken from~\cite{Billard:2013qya}.
  }
  \label{fig:dm_observables}
\end{figure}

We further investigate the VEV configurations that allow for an SFOEWPT.
Figure~\ref{fig:vevevo_cpv_ft} illustrates the temperature dependence of all VEV directions, as well as the electroweak VEV,
\begin{align}
  v(T) =
  \sqrt{\overline{\omega}_1^2(T)+\overline{\omega}_2^2(T)+\overline{\omega}_{\text{CP}}^2(T)+
  \overline{\omega}_{\text{CB}}^2(T)}\,\label{eq:EWVEV},
\end{align}
for two SFOEWPT points, one for each of the two categories of VEV-configuration temperature evolution.
Most sample points show a VEV evolution as a function of the temperature as illustrated in Fig.~\ref{fig:vevevo_cpv_ft} (left), where only the SM-like VEV $\overline{\omega}_1$ participates in the EWPT and all dark VEVs are zero over the whole temperature range.
However, we also find viable SFOEWPT points that show a similar VEV-configuration evolution as shown in Fig.~\ref{fig:vevevo_cpv_ft} (right), where all dark VEVs (except the charge-breaking dark VEV) are non-zero at $T=T_c$.
All points of the sample are found to have a numerically zero charge-breaking dark VEV over the whole temperature range as required for a massless photon.
All (non-charge breaking) dark VEVs participate in the SFOEWPT for points of the second category, moreover, they are non-zero for a finite range in finite temperature.
A non-zero $\overline{\omega}_\text{CP}$ corresponds to the generation of \textit{spontaneous} CP violation.
In Fig.~\ref{fig:spon_cpv_ft} we show the absolute value at the global minimum at $T=T_c$ of the CP-violating VEV direction $|\overline{\omega}_\text{CP}|$ versus the ratio $|\Im(A)/\Re(A)|$ as a relative measure for the (zero-temperature) dark explicit CP violation.
We note, that we have $|\overline{\omega}_\text{CP}|>0$ only for $\Im(A)\neq 0$, but there is no clear correlation between the size of $|\overline{\omega}_\text{CP}|$ and $\Im(A)\neq 0$ or $\xi_c$.
Non-zero dark VEVs in a finite range in finite temperature in addition lead to spontaneous breaking of the $\mathbb{Z}_2$-symmetry.
At finite temperature, dark charge therefore is not conserved, and particles that are dark at $T=\SI{0}{GeV}$ can now mix with particles stemming from the first doublet.
This opens a promising portal for the transfer of non-standard CP violation to the SM-like Higgs couplings to fermions at finite temperature, which is not constrained by the EDM measurements at zero temperature.\s
Finally, we present our parameter sample in the parameter space spanned by the relic density and DM mass $m_{h_1}$ in Fig.~\ref{fig:dm_observables} (left).
All SFOEWPT points provide a viable, here underabundant particle DM candidate, compatible with the measured relic density.
Taking this underabundance into account via a rescaling factor $f_{\chi\chi}=\frac{\Omega_\text{prod}h^2}{\Omega_\text{obs}h^2}$, we display the effective spin-independent (SI) nucleon DM cross section versus the DM mass $m_{h_1}$ in Fig.~\ref{fig:dm_observables} (right).
As already required by {\tt ScannerS}, all points lie below the {\tt XENON1T} exclusion limit~\cite{XENON:2018voc}.
We find the majority of the SFOEWPT points to lie above the neutrino floor and above the expected sensitivity of the {\tt XENONnT} experiment~\cite{XENON:2020kmp}.
Consequently, future DM direct detection experiments will allow us to test a large fraction of the parameter space of the model that allows for an SFOEWPT.

\section{Conclusion}\label{sec:conclusion}
`CP in the Dark' proves itself to be a promising BSM candidate in the context of explaining DM and the BAU through EWBG. It provides a DM candidate as well as viable SFOEWPT points in a large range of the allowed parameter space.
In addition to having explicit CP violation in its dark sector, it allows for spontaneous CP violation at finite temperature.
Future increases in the precision of collider and DM direct detection experiments will be able to test a large fraction of the parameter space of the model.

\acknowledgments
MM acknowledges financial support by the Deutsche Forschungsgemeinschaft (DFG, German Research Foundation) under grant 396021762 - TRR 257.
The work of JM was supported by the BMBF-Project 05H18VKCC1.
We thank Philipp Basler, Duarte Azevedo, G\"unter Quast, Jonas Wittbrodt and Michael Spira for fruitful discussions.

\appendix

\end{document}